\documentclass[12pt]{article}
\usepackage[english]{babel}
\usepackage[dvips]{graphicx}
\usepackage{mathtext}
\usepackage{amsmath}
\usepackage{amssymb}

\begin{document}

\title{\bf Tachyonic fields in cosmology}
\author{Bohdan Novosyadlyj  \\
\textit{Astronomical observatory of Ivan Franko National University of Lviv,}\\
\textit{Kyryla i Methodia str., 8, Lviv, 79005, Ukraine}}
\maketitle

{\small
The possibility of explanation of accelerated expansion of the Universe by tachyonic scalar fields which homoge\-neously fill the world is discussed. The dependences of potential and kinetic term on scale factor are deduced for the case of quintessential and phantom dark energy with generalized linear barotropic equation of state. The possibility to distinguish the tachyonic scalar field as dark energy from other scalar field models, especially from classical scalar field, is analyzed.

\medskip

}

\section*{Introduction}
The concept of tachyons \cite{Bilaniuk1962} as consistent physical theory of particles that move with superluminal velocities has already more than 50 years, however up to now we have no experimental proof of their existence or irrefutable arguments against them. Nevertheless, the hypothesis proved to be extremely fruitful for physics in general, not only for development of the theory of relativity. Tachyonic modes of oscillations of branes and strings have been discussed already for a long time in boson and superstring theories of fundamental interactions (see, for example, \cite{Garousi2000,Sen2002}). Tachyonic fields are also considered as possible candidates for physical essence that caused exponential expansion and physical processes in very early Universe \cite{Garousi2004} and accelerated expansion of Universe in our epoch \cite{Padmanabhan2002,Gibbons2002,Frolov2002,Bagla2003,Abramo2003,Gibbons2003,Abramo2004,Gorini2004,Sen2005,Calcagni2006}.
From all physical theories, which use the term "tachyon", coined by  G. Feinberg in 1967 \cite{Feinberg1967}, only the latter have been  proved experimentally by two science teams in 1998 \cite{Perlmutter1998,Riess1998,Schmidt1998}, for what their supervisors were awarded the Nobel Prize in physics in 2011.
This stimulates the scientists to use the data of observational cosmology for search the hints of possible existence of something tachyonic in our world, for example, a cosmological scalar field. It appears that such field has a number of interesting physical features and cosmological consequences, study of which expands our physical outlook and enriches world's treasury of scientific work, such as our guesses about possible forms of matter in our Universe or in other hypothetical worlds.

In this paper, based on the author's lecture on the same subject given on physical seminar of Shevchenko Scientific Society devoted to the 50th anniversary of concept of tachyons, the main features of cosmological tachyon scalar field are reviewed, as well as the possibility of determination of their parameters based on modern and expected observational cosmological data is discussed.

\section{Tachyonic scalar field as dark energy accelerating expansion of the Universe}

We assume that the Universe is homogeneous and isotropic, the metric of 4-space is Friedmann-Robertson-Walker (FRW) one,
\begin{equation}\label{ds_frw}
ds^2=g_{ij} dx^i dx^j = a^2(\eta)\left(d\eta^2-dr^2-\chi^2(r)(d\vartheta^2+\sin^2\vartheta d\varphi^2)\right],
\end{equation}
where $\eta$ is conformal time defined as $dt=a(\eta)d\eta$ (hereafter we put $c=1$, hence time variable $t\equiv x^0$ has dimension of length), and the factor $a(\eta)$ is radius of 3-sphere in the case of 3-space of positive curvature, radius of 3-pseudosphere in the case of 3-space of negative curvature or scale factor in the case of 3-space of zero curvature (flat or Euclidean 3-space). In the latter case it is convenient to norm it to 1 in current epoch, $a(\eta_0)=1$. Hereafter the Latin indexes $i,\,j,\,...$ run the values 0, 1, 2, 3 and Greek ones $\nu,\,\mu,\,...$ -- 1, 2, 3. The function $\chi(r)$ depends on curvature of 3-space $K$,
\begin{equation}
\label{chir}
\chi(r) = 
\begin{cases} 
  \frac{1}{\sqrt{K}} \sin\sqrt{K}r,  & K>0. \\
  r,      & K=0. \\
  \frac{1}{\sqrt{|K|}}\sinh\sqrt{|K|}r, & K<0.  
  \end{cases}
\end{equation}
We assume also that the Universe is filled with non-relativistic (cold dark matter and baryons) and relativistic particles (thermal electromagnetic radiation and massless neutrinos) and also with dark energy, which interacts with other components only gravitationally (minimal coupling), so the dynamics is completely described by Einstein equation
\begin{eqnarray}
R_{ij}-{\frac{1}{2}}g_{ij}R=8\pi G \left(T_{ij}^{(m)}+T_{ij}^{(r)}+T_{ij}^{(de)}\right),
\label{E_eq}
\end{eqnarray}
where $R_{ij}$ is Ricci tensor and $T_{ij}^{(m)}$, $T_{ij}^{(r)}$, $T_{ij}^{(de)}$ are energy-momentum tensors of non-relativistic $m$, relativistic $r$ matter and dark energy $de$ respectively.

In the case of only gravitational coupling between components each of them should satisfy the differential energy-momentum conservation law separately:
\begin{eqnarray}
T^{i\;\;(n)}_{j\;;i}=0.\label{conseq_bckgr}
\end{eqnarray}
Hereafter ``;'' denotes covariant derivative with respect to the coordinate with given index in space with metric (\ref{ds_frw}) and ($n$) means $m$, $r$ or $de$.
For perfect fluid with density $\rho_{(n)}$ and pressure $p_{(n)}$, connected by equation of state $p_{(n)}=w_{(n)}\rho_{(n)}$, equation (\ref{conseq_bckgr}) leads to:
\begin{eqnarray}
\dot{\rho}_{(n)}=-3\frac{\dot a}{a} \rho_{(n)}(1+w_{(n)})\label{rho'},
\end{eqnarray}
hereafter dot is the derivative with respect to the conformal time: ``$\dot{\;\;}$''$\equiv d/d\eta$. For non-relativistic matter 
$w_{m}=0$ and $\rho_{m}=\rho_{m}^{(0)}a^{-3}$, for relativistic one $w_{r}=1/3$ and $\rho_{r}=\rho_{r}^{(0)}a^{-4}$. Hereafter ``0'' denotes the current value of physical quantity.

We model the dark energy by a scalar field which violates the strong energy condition, homogeneously fills the Universe and interacts with other components only gravitationally. Important problem of modern cosmology is reduction of the number of possible models of such fields by comparison of theoretical predictions with corresponding observational data. Myriads of models of cosmological scalar fields can be classified 
by the region of values of EoS parameter as quintessence ($-1<w_{de}<-1/3$) and phantom ($w_{de}<-1$), by the time dependence of EoS parameter as freezing ($\dot{w}_{de}<0$) and thawing ($\dot{w}_{de}>0$), by Lagrangian type as classical, tachyon, quintom, K-essence and so on. Among them, tachyon scalar fields belong to the most perspective ones, since they are connected with superstring and early Universe theories.

Scalar field $\phi(x^i)$ with Dirac-Born-Infeld Lagrangian 
\begin{equation}
  \mathcal{L}=-U(\phi)\sqrt{1-2X},\label{L_tach}
\end{equation}
where $X\equiv\phi_{;i}\phi^{;i}/2$ is kinetic term and $U$ is potential, is called tachyonic. In homogeneous isotropic Universe with FRW metric the field is homogeneous $\phi(x^0)$ and its effective energy density and pressure are defined by values of $X$ and $U$:
\begin{equation}
\rho_{de}\equiv2X\mathcal{L}_{,X}-\mathcal{L}=\frac{U(\phi)}{\sqrt{1-2X}}, \quad
p_{de}\equiv\mathcal{L}=-U(\phi)\sqrt{1-2X}. \label{rho_XU_tach}
\end{equation}
 The ratio of these two values $w_{de}\equiv p_{de}/\rho_{de}$, often called the EoS parameter similarly to the matter components, is defined by the value of kinetic term $X$ only:
\begin{equation}
  w_{de}=2X-1.
  \label{w_XU_tach}
\end{equation}
These quantities can be evaluated by solving the equation of motion (or Euler-Lagrange equation) for the field 
\begin{equation}
  \left(\ddot{\phi}+2aH\dot{\phi}\right)\mathcal{L}_{,X}-a^2\frac{\partial{U}}{\partial{\phi}}\mathcal{L}_{,U}+\frac
  {\ddot{\phi}\dot{\phi}^2-aH\dot{\phi}^3}{a^2}\mathcal{L}_{,XX}+\frac{\partial{U}}{\partial{\phi}}\dot{\phi}^2\mathcal{L}_{,XU}=0.
  \label{gen_KG}
\end{equation}
which in our case (\ref{L_tach}) takes the form \cite{Sergijenko2009b}:
\begin{equation}
\frac{\ddot{\phi}-aH\dot{\phi}}{1-\left(\dot{\phi}/a\right)^2}+3aH\dot{\phi}+a^2\frac{1}{U}\frac{dU}{d\phi}=0,
\label{KG}
\end{equation}
where $H\equiv\dot{a}/a^2$ is Hubble parameter (on the hypersurfaces of constant time). For the solution the explicit functional dependences of potential $U$ on field variable $\phi$ and of Hubble parameter on time must be known. The latter can be found from Friedmann equation as one of Einstein equations (\ref{E_eq}) in the world with metric (\ref{ds_frw})
\begin{equation}
H =  H_0\sqrt{\Omega_r \frac{a_0^4}{a^4} +  \Omega_m\frac{a_0^3}{a^3}+\Omega_K\frac{a_0^2}{a^2} + \Omega_{de}f(a)},
\label{H}
\end{equation}
where the dimensionless constants
\begin{equation}
\Omega_m \equiv \left(\frac{\rho_m}{\rho_{cr}}\right)_{\eta_0}, \quad \Omega_r \equiv \left(\frac{\rho_r}{\rho_{cr}}\right)_{\eta_0}, \quad
\Omega_{de} \equiv \left(\frac{\rho_{de}}{\rho_{cr}}\right)_{\eta_0}, \quad \Omega_K \equiv\left(\frac{-K}{(\dot{a}/a)^2}\right)_{\eta_0}.
\label{def1}
\end{equation}
are the densities of the components of the Universe in current epoch in units of critical density and curvature in the units of Friedmann radius of the world. Among these dimensionless constants the density of relativistic component is established most precisely:
\begin{equation}
  \label{Om_r}
  \Omega_{r}= 4.17\cdot 10^{-5}\left(\frac{1+\rho_{\nu}/\rho_{\gamma}}{1.6813}\right)\left(\frac{T_0}{2.726}\right)^4
  \approx 4.17\cdot 10^{-5}h^{-2}.
\end{equation}
Function $f(a)$ describes the dynamics of changing of energy density of scalar field, so that $f(a_0)=1$. We can find it with (\ref{rho_XU_tach}) and $\phi(a)$ as the solution of (\ref{KG}).
Therefore, in such approach the evolution of the field with given functional dependence of $U(\phi)$ and the dynamics of expansion of the Universe with given parameters $H_0$, $\Omega_m$, $\Omega_K$ and $\Omega_{de}$ can be described by the consistent solution of equations (\ref{KG})-(\ref{H}) and (\ref{rho_XU_tach}). Another, much more productive approach consists in solving of the simple differential linear homogeneous 1st order equation (\ref{rho'}) by specifying function $w_{de}(\eta)$ instead of solving quasilinear differential 2nd order equation (\ref{KG}) and with help of equations (\ref{rho_XU_tach}) reconstructing $\phi(\eta)$, $U(\eta)$ and $U(\phi)$. It is convenient, however, to use dependence of the variables on $a$ and derivative in respect to it instead of dependence on conformal time $\eta$ and its derivatives ($(^.)\equiv d/d\eta=a^2Hd/da$). Equation (\ref{rho'}) in this case becomes
\begin{eqnarray}
\frac{d\ln{\rho_{de}}}{d\ln{a}}=-3(1+w_{de})\label{rho''},
\end{eqnarray}
which immediately gives function $f(a)$ in (\ref{H}) when $w_{de}(a)$ is given,
\begin{eqnarray}
f(a)=(a_0/a)^{3(1+\tilde{w}_{de})}, \quad \tilde{w}_{de}=\frac{1}{\ln{(a/a_0)}}\int_{a_0}^{a}{w_{de}d\ln{a}} \label{f_a},
\end{eqnarray}
Field variable $\phi(a)$, potential $U(a)$ and kinetic term $X(a)$ are expressed via them as follows:
\begin{equation}
\label{U_tach}
\begin{aligned}
  \phi(a)-\phi_0 & =\pm\int_1^a\frac{da'\sqrt{1+w_{de}(a')}}{a'H(a')},\\
  U(a) & =\rho_{de}(a)\sqrt{-w_{de}(a)}, \\
  X(a) & =\frac{1+w_{de}(a)}{2}.
\end{aligned}
\end{equation}
We can see that field variable $\phi(a)$ is real if $w_{de}(a)>-1$. Kinetic term is always positive in this case. With positively-defined energy density of tachyon field the potential is positive if $w_{de}(a)<0$. With this condition the energy density $\rho_{de}(a)$ and pressure $p_{de}(a)$ are always real. On the other hand,
$\left(^{\alpha}_{\alpha}\right)$-Einstein equations in world with metric ($\ref{ds_frw}$) imply the Friedmann equations with second order derivative of the radius of the world with respect to time,
\begin{equation}
  q = \frac{H_0^2}{H^2}\left[\Omega_r \frac{a_0^4}{a^4} +
    \frac{1}{2}\Omega_m\frac{a_0^3}{a^3}
    +\frac{1}{2}(1+3w_{de})\Omega_{de}f(a) \right], \label{q}
\end{equation}
where $q\equiv-\left(a\ddot{a}/\dot{a}^2- 1\right)$ is usually called the deceleration parameter. The accelerated expansion of the Universe in the current epoch\footnote{Here and after we omit component with $\Omega_r$, because in epoch $a\sim 1$ $\rho_r\ll \rho_m$.} ($q_0<0$), detected from luminosity distance - redshift relation data for SNe Ia, can be obtained when EoS parameter satisfies the condition: 
\begin{equation}
   w_{de}<-\frac{1}{3}-\frac{\Omega_m}{\Omega_{de}}. \label{ac}
\end{equation}
Thus, real scalar field $\phi$ with tachyon potential (\ref{L_tach}) will have real positive values of potential, kinetic term and energy density and will cause the accelerated expansion of the Universe, if its EoS parameter $w_{de}$ is in the range
\begin{equation}
  -1<w_{de}<-\frac{1}{3}-\frac{\Omega_m}{\Omega_{de}}. \label{cond_w}
\end{equation}
For the field values this condition has the following form:
\begin{equation}
  \text{ a)}\,\, 0<X^{(0)}<\frac{1}{3}-\frac{\Omega_m}{2\Omega_{de}}, \quad
  {\rm b)}\,\, U^{(0)}\frac{1+3X^{(0)}}{\sqrt{1-2X^{(0)}}}>\rho_m^{(0)}/2.
  \label{cond_U}
\end{equation}
Important characteristic of field is also the effective speed of propagation of perturbations (effective sound speed), square of which is determined by the Lagrangian as follows
\begin{equation}
  c^2_s\equiv\frac{\delta p}{\delta \rho}=\frac{\mathcal{L}_{,X}}{\mathcal{L}_{,X}+2X\mathcal{L}_{,XX}}.
  \label{c_s2}
\end{equation}
In the case of tachyon field with Lagrangian (\ref{L_tach}) $c^2_s=-w_{de}$ and  taking into account (\ref{cond_w}) the value of square of the effective sound speed is in the range $1/3<c^2_s<1$, which ensures the gravitational stability of this component of the Universe. Thus, tachyonic scalar field can be quintessential dark energy, as the conditions (\ref{cond_w}) and (\ref{cond_U}) are consistent, the field values and energy density are real and speed of propagation of the perturbations is less than speed of light. Such tachyonic scalar field has almost nothing common with superluminal tachyons.
\begin{figure}[tbp]
  \centering
  \includegraphics[width=.47\textwidth]{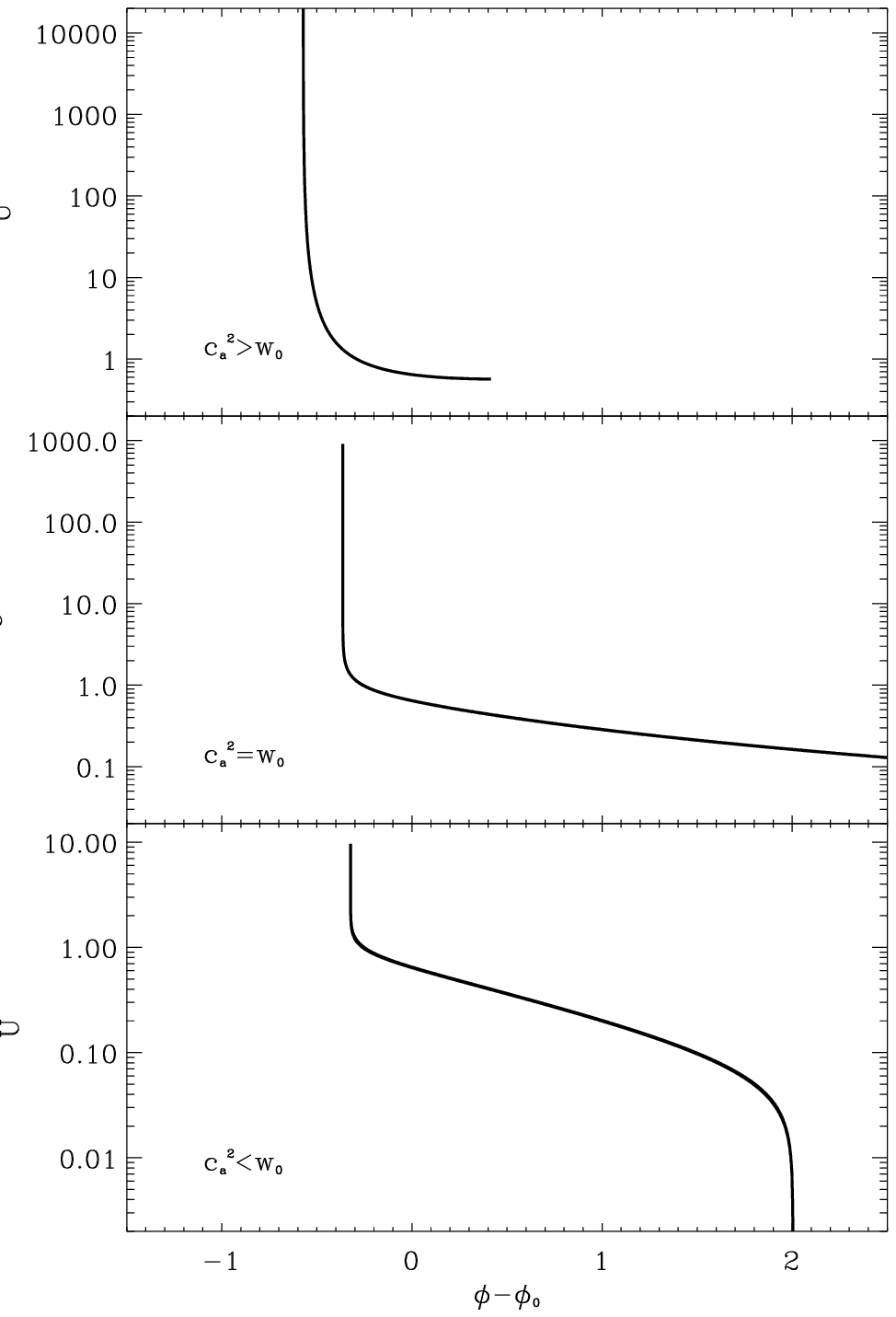}
  \includegraphics[width=.47\textwidth]{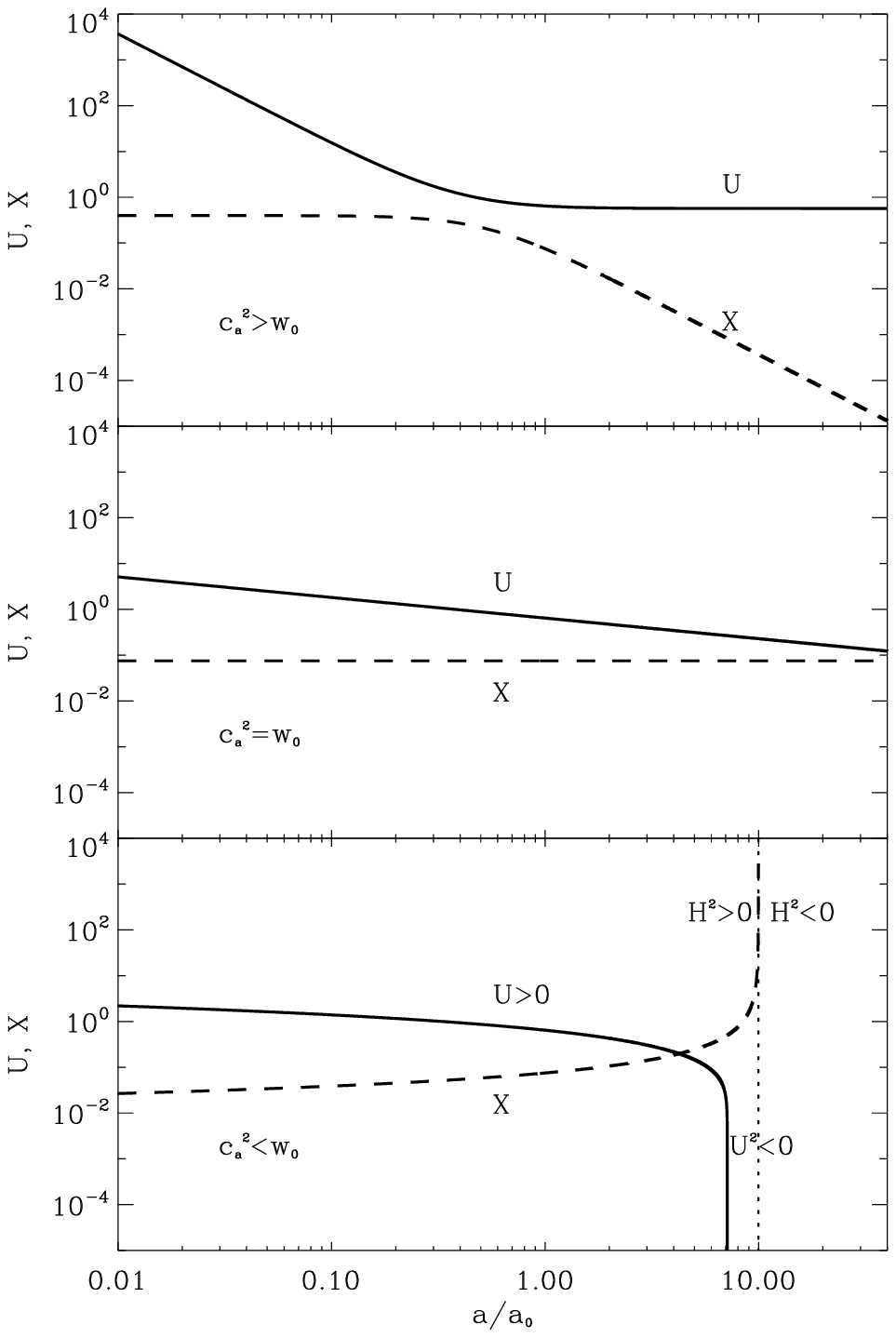}
  \caption{Potentials $U(\phi-\phi_0)$ (left) and dependence of potentials and kinetic terms on $a$ (right) for tachyon scalar fields with decreasing, constant and increasing EoS parameter (from top to bottom). Potentials and kinetic terms are represented in units of critical density in the current epoch, $3c^2H_0^2/8\pi G$, field variable $\phi$ is in units $\sqrt{3c^2/8\pi G}$. Current epoch ($a/a_0=1$) in right panels corresponds to the value $\phi-\phi_0=0$ and field evolves from left to right.}
  \label{U_X_tach}
\end{figure}

Let us specify the field by defining $w_{de}$ as follows. EoS parameter of dark energy $w_{de}$ and its adiabatic speed of sound $c_{a\;(de)}^2\equiv\dot{p}_{de}/\dot{\rho}_{de}$ are connected by ordinary differential equation
\begin{eqnarray}
 w_{de}'=\frac{3}{a}(1+w_{de})(w_{de}-c_a^2),\label{w'}
\end{eqnarray}
where a prime (') denotes the derivative with respect to $a$. It is necessary to note that $c_a^2$ is not actually the adiabatic sound speed in dark energy. It is only one of the phenomenological parameters, which describe dark energy, but we call it ``adiabatic sound speed'' similarly to the thermodynamical variable with the same definition. It is easy to see that the derivative of EoS parameter with respect to scale factor is negative for $w_{de}<c_a^2$ and positive for $w_{de}>c_a^2$. In the first case in process of evolution of the Universe the dark energy acquires the ability to accelerate the expansion, in the second one it loses such ability. In general case $c_a^2$ can be function of time and equation (\ref{w'}) is the differential Riccati equation. However hereafter we assume that it is constant: $c_a^2=const$. Such assumption is the simplest one and it allows to obtain the analytical solutions. The time derivative of pressure $p_{de}(\eta)$ is then proportional to the time derivative of density 
$\rho_{de}(\eta)$. The integral form of this condition is generalized linear barotropic equation of state
\begin{eqnarray}
 p_{de}=c_a^2\rho_{de}+C,\label{beos}
\end{eqnarray}
where $C$ is constant. Cosmological scenarios of evolution of the Universe filled with fluid with such EoS (also called ``wet dark fluid'') were analyzed in \cite{Babichev2005a,Holman2004}.
Solution of differential equation (\ref{w'}) with $c_a^2=const$ is:
\begin{eqnarray}
 w_{de}(a)=\frac{(1+c^2_a)(1+w_0)}{1+w_0-(w_0-c^2_a)(a/a_0)^{3(1+c^2_a)}}-1,\label{w}
\end{eqnarray}
where integration constant $w_0$ is taken to be equal to the current value of $w_{de}$, when $a=a_0$. It is easy to see that definition of $w_{de}$ and expressions (\ref{beos})-(\ref{w}) lead to $C=\rho_{de}^{(0)}(w_0-c_a^2)$, where $\rho_{de}^{(0)}$ is current value of the energy density of dark energy. Thus, two quantities $w_0$ and $c_a^2$ define EoS parameter $w_{de}$ at any redshift $z=a_0/a-1$. From (\ref{w}) it follows that $c_a^2$ corresponds to EoS parameter at the beginning of the expansion of the Universe ($w_{init}\equiv w_{de}(0)=c_a^2,\,a=0,\,z=\infty$). Differential equation (\ref{rho'}) with $w_{de}$ in form of (\ref{w}) also has the analytical solution
\begin{eqnarray}
\rho_{de}=\rho_{de}^{(0)}\frac{(1+w_0)(a/a_0)^{-3(1+c_a^2)}+c_a^2-w_0}{1+c_a^2},\label{rho}
\end{eqnarray}
which yields the analytical expression for $f(a)\equiv \rho_{de}/\rho_{de}^{(0)}$.

The dependence of potential $U$ on field variable $\phi$ and scale factor $a/a_0$ is shown for tachyon scalar field with decreasing, constant and increasing EoS parameter (\ref{w}) in fig.~\ref{U_X_tach}. Such field ensures the accelerated expansion of the Universe with deceleration parameter in current epoch $q_0\approx-0.5$ (see \cite{Sergijenko2009a,Novosyadlyj2010} for details), which follows from the data on luminosity-distance relation for supernovae type Ia. 

One can note that accelerated expansion of the Universe is caused by slow rolling of the tachyon field potential to the minimum. The dynamics of expansion of the Universe at all stages is determined only by $\Omega_{de}$ and $w_{de}$ and does not depend on the Lagrangian of scalar field, as it follows from (\ref{H}) and (\ref{q}). We have shown this for scalar field with classical and tachyon Lagrangians \cite{Novosyadlyj2010,Novosyadlyj2011}) and have concluded that other tests, sensitive to physical features of different fields, should be found.

In research by different authors in last few years there are more and more arguments that dark energy has phantom nature (see \cite{Novosyadlyj2012} and citation in it), so let us analyze the possibility of modeling of such dark energy by tachyonic scalar field.

\begin{figure}[tb]
  \centering
  \includegraphics[width=.47\textwidth]{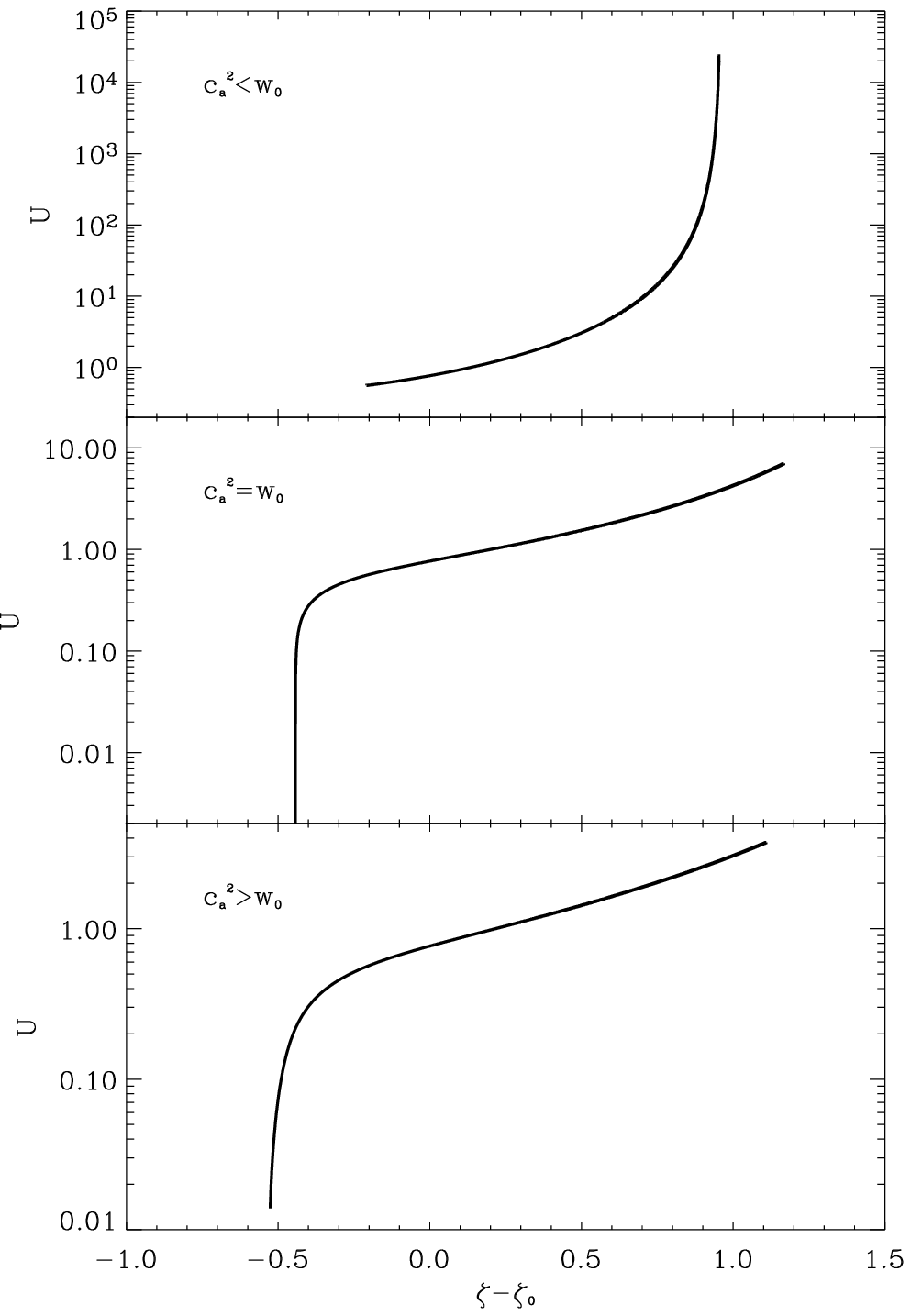}
  \includegraphics[width=.47\textwidth]{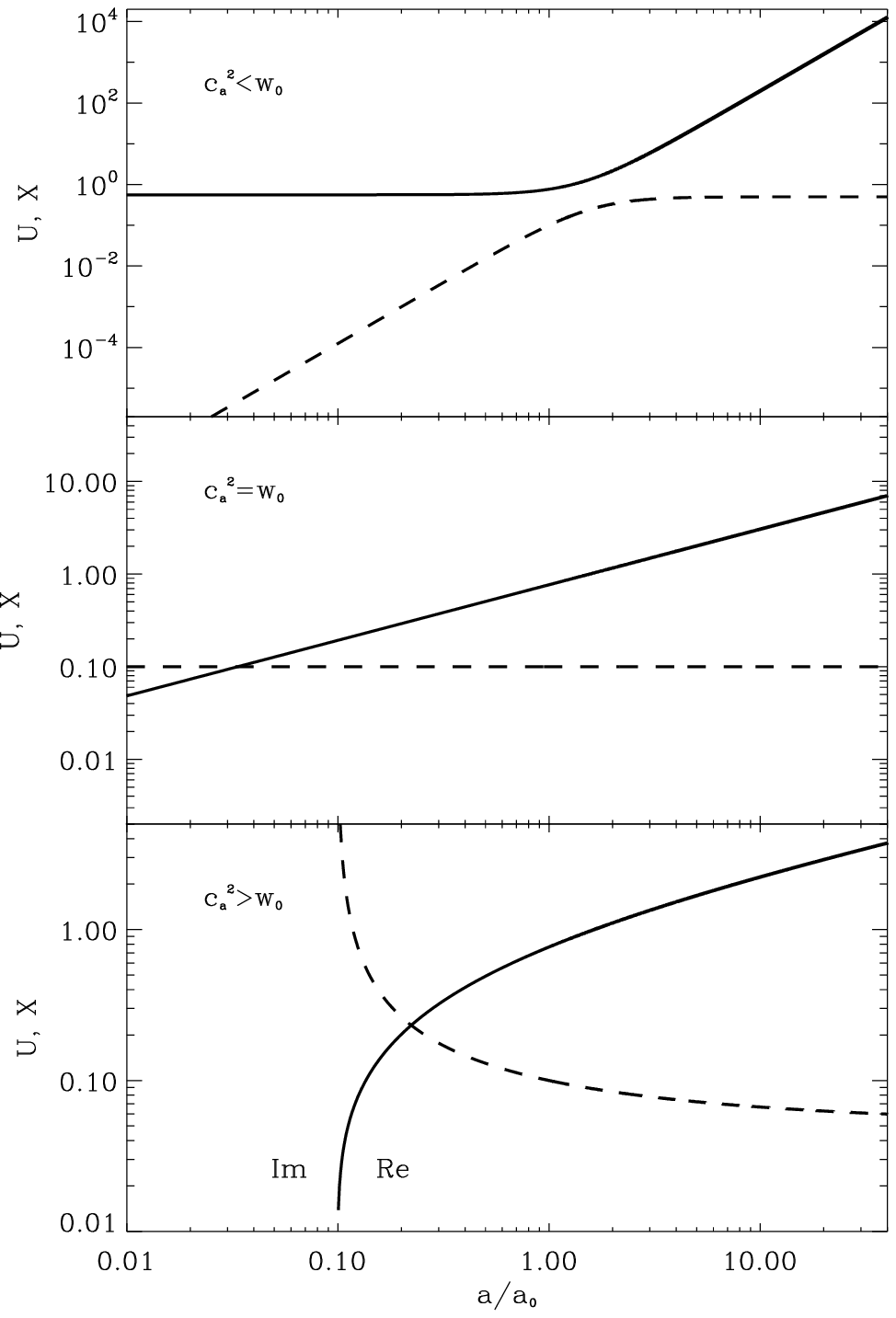}
  \caption{Potentials $\tilde{U}(\xi-\xi_0)$ (left) and dependence of potentials and kinetic terms on $a/a_0$ (right) for phantom tachyonic field with decreasing, constant and increasing EoS parameter (from top to bottom). The units are the same as in fig.~\ref{U_X_tach}. }
\label{U_X_ph2}
\end{figure}
 
From the expressions (\ref{U_tach}) it follows that scalar field with the Lagrangian (\ref{L_tach}) can be phantom ($w_{de}<-1$) when field variable $\phi$ is imaginary and kinetic term is negative. But let us change the sign of kinetic term in Dirac-Born-Infeld Lagrangian, similarly to obtaining the phantom field from classical one in \cite{Caldwell2002}:
\begin{equation}
 \mathcal{L}=-\tilde{U}(\xi)\sqrt{1+2\tilde{X}},\label{L_ph2}
\end{equation}
where $\tilde{U}$ and $\tilde{X}$ are potential and kinetic term of phantom tachyonic field. The
density, pressure and EoS parameter in this case are as follows:
\begin{equation}
  \rho_{de} = \frac{\tilde{U}(\xi)}{\sqrt{1+2\tilde{X}}}, \qquad p_{de}=-\tilde{U}(\xi)\sqrt{1+2\tilde{X}},   \qquad
  w_{de} = -2\tilde{X}-1.\label{w_XU_ph2}
\end{equation}
Hence, with any positive $\tilde{X}$ and $\tilde{U}$ the energy density is real positive number, the pressure and EoS parameter are negative numbers, besides that, $w_{de}\le-1$ when $\tilde{X}\ge0$ in any case. For the explanation of accelerated expansion of the Universe in current epoch the values $\tilde{X}$ and $\tilde{U}$ must satisfy next conditions:
\begin{equation}
  \text{ a)}\,\, \tilde{X}^{(0)},\,\tilde{U}^{(0)}>0, \quad  \text{ b)}\,\, \tilde{U}^{(0)}\frac{1+3\tilde{X}^{(0)}}{\sqrt{1+2\tilde{X}^{(0)}}}>\rho_m^{(0)}/2.
\end{equation}
In the case of phantom field (\ref{L_ph2}) its field variable, potential and kinetic term are determined by energy density and EoS parameter as follows:
\begin{equation}
\label{U_ph2}
\begin{aligned}
  & \xi(a)-\xi_0=\pm\int_1^a\frac{da'\sqrt{-(1+w_{de}(a'))}}{a'H(a')},\\
  & \tilde{U}(a)=\rho_{de}(a)\sqrt{-w_{de}(a)}, \\
  & \tilde{X}(a)=-\frac{1+w_{de}(a)}{2}.
\end{aligned}
\end{equation}

The potential $\tilde{U}(\xi-\xi_0)$, the evolution of potential and kinetic term in the models with decreasing, constant and increasing EoS parameter are shown in fig.~\ref{U_X_ph2}. As in the case of the classical field with inverse sign of the kinetic term in Lagrangian, potential of phantom field (\ref{L_ph2}) during the process of expansion of the Universe rolls up, this causes the acceleration $\ddot{a}>0$ in the current epoch. Another fascinating feature of this phantom field consists in the fact that its effective sound speed, which is equal to $-w_{de}$ according to (\ref{c_s2}), and is larger than speed of light ($c_s^2>1$). But since the field interacts only gravitationally with all other components, we suppose that this fact does not lead to violation of the causality principle. Thus, such field is interesting yet from the point of view of the possibility of superluminal  propagation of perturbations, which, however, needs separate detailed analysis. Similar analysis was made for k-essential 
fields in \cite{Babichev2008}, in which the absence of violation of the causality principle for minimally coupled dark energy, in which the effective speed of propagation of perturbations is larger than speed of light, is noted.  

\section{Gravitational instability of tachyonic scalar field and possibility of distinguishing}

Thus, tachyonic scalar field is indistinguishable from classical one through its influence on dynamics of the expansion of the homogeneous isotropic Universe, though the dynamics of changes of the field values is different, this one can see comparing fig.~\ref{U_X_tach} here with fig.~3 in \cite{Novosyadlyj2010} and fig.~\ref{U_X_ph2} here and fig.~2 in \cite{Novosyadlyj2012}. That is why we consider evolution of the scalar perturbations in many-component Universe. The evolution of magnitudes of perturbations depends on effective sound speed of each component and gravitational influence of the perturbations of the dark energy on the dark matter perturbations. Since the effective sound speed of classical field is constant and equals to the speed of light while for tachyonic field it equals to $-w_{de}(a)$, we can expect special features in the formation of the large scale structure of the Universe and possibility of distinguishing of the tachyonic and classical scalar fields.

Let us consider the scalar perturbations in synchronous gauge with flat 3-space world metric:
\begin{eqnarray}
ds^2=g_{ij} dx^i dx^j =a^2(\eta)\left[d\eta^2-(\delta_{\alpha\beta}+h_{\alpha\beta}) dx^{\alpha}dx^{\beta}\right].\label{ds}
\end{eqnarray}
As a rule the scale factor in cosmological models with flat 3-space is normalized to 1, $a_0=1$. Scalar perturbations of metric $h_{\alpha\beta}$ can be decomposed into trace $h\equiv h_{\alpha}^{\alpha}$ and traceless
$\tilde{h}_{\alpha\beta}$ components as $h_{\alpha\beta}=h\delta_{\alpha\beta}/3+\tilde{h}_{\alpha\beta}$. For small perturbations ($h\ll1$) all  equations can be linearized with respect to the perturbed variables. In the multi-component model of the Universe each component is moving with small peculiar four-velocity $\delta u^{\alpha}\equiv dx^{\alpha}/ds\ll1$, which is determined by properties of the given component (density, pressure, entropy, sound speed and so on) and $h$. In the case of scalar mode of perturbations the spatial part of the four-velocity $\delta u^{\alpha}$ can be expressed as gradient of a scalar function $V(\eta,\textbf{x})$: $\delta u^{i} = g^{ij}V_{,j}$ for each component. As the cold dark matter (CDM) is assumed to be perfect fluid with zero pressure interacting with other components only gravitationally, synchronous gauge is usually defined as comoving to particles of CDM. In the linear perturbation theory it is convenient to perform the Fourier transformation of all spatially-
dependent variables and use the equations for corresponding Fourier amplitudes.
\begin{figure}[tb]
  \centering
  \includegraphics[width=.47\textwidth]{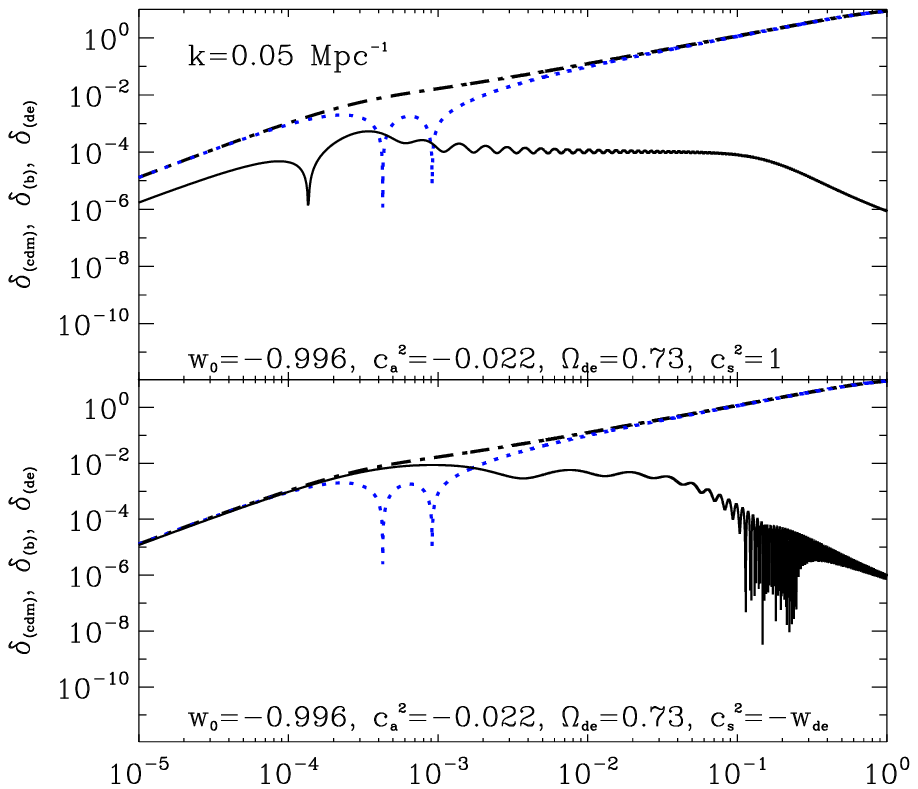}
  \includegraphics[width=.47\textwidth]{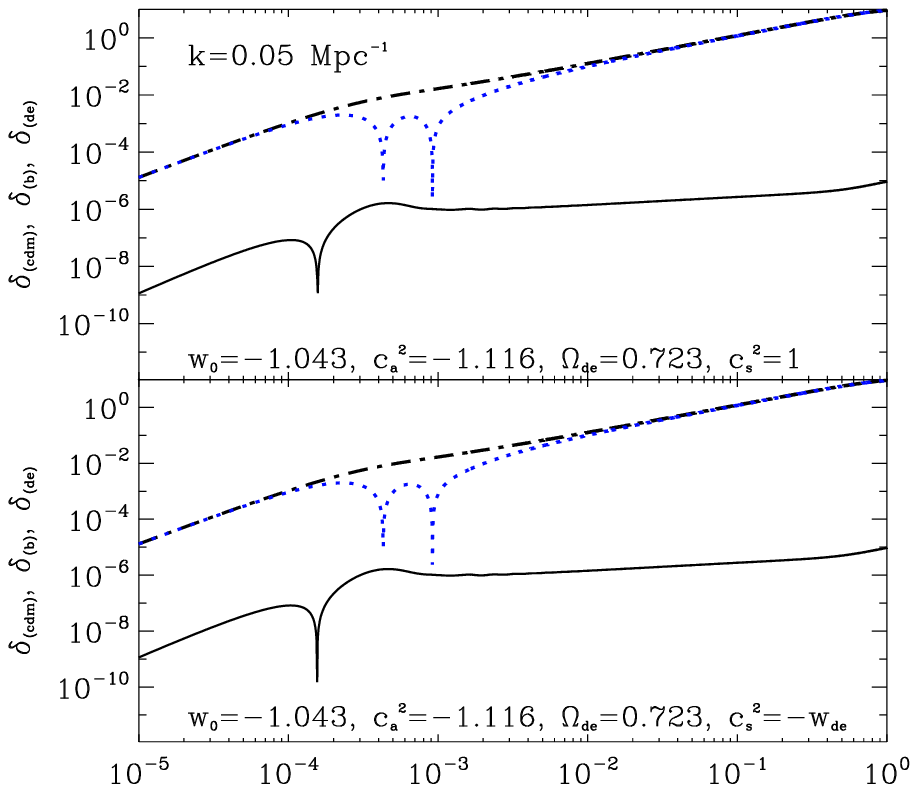}
  \caption{Evolution of the cosmological perturbations of density of dark energy (solid line), dark matter (dashed line) and baryonic component (doted line) in models with classical ($\mathcal{L}=\pm X-U$, upper panels) and tachyonic ($\mathcal{L}=-U(\phi)\sqrt{1\mp 2X}$, lower panels) quintessential (left, upper sign in the Lagrangians) and phantom (right, lower sign in the Lagrangians) scalar fields as a dark energy.}
\label{ddeb}
\end{figure}

The differential energy-momentum conservation law $\delta T^{i\;\;(de)}_{j\;;i}=0$ for the perturbations in space with metric (\ref{ds}) leads us to the following equations for the evolution of perturbations of density and velocity of dark energy in synchronous gauge:
\begin{eqnarray}
&&\dot{\delta}_{de}+3(c_s^2-w_{de})aH\delta_{de}+(1+w_{de})\frac{\dot{h}}{2}
+(1+w_{de})\left[k+9a^2H^2\frac{c_s^2-c_a^2}{k}\right]V^{(de)}=0, \label{d_de}\\
&&\dot{V}_{de}+aH(1-3c_s^2)V_{de}-\frac{c_s^2k}{1+w_{de}}\delta_{de}=0.\label{V_de}
\end{eqnarray}
Equations for metric perturbations as well as perturbations of densities and velocities of relativistic and non-relativistic components (photons, massless neutrinos, baryons and cold dark matter) are presented in \cite{Ma1995}. For the analysis of the evolution of cosmological perturbations it is important to set properly the initial conditions. It is known that large scale structure of the Universe was formed from small adiabatic perturbations generated in early Universe. Therefore, for all components of the Universe, except for dark energy, the adiabatic initial conditions are taken from \cite{Ma1995}. Initial conditions for the perturbations of dark energy are obtained from asymptotic solutions (\ref{d_de})-(\ref{V_de}) for $k\eta\ll1$ (perturbations with scale larger than particle horizon) in early radiation-dominated epoch:
\begin{eqnarray}
&&\delta_{de}^{\;init} = -\frac{(4-3c_s^2)(1+w_{de})}{8+6c_s^2-12w_{de}+9c_s^2(w_{de}-c_a^2)} h_{init},\label{d_de_init}\\
&&V_{de}^{\;init} = -\frac{c_s^2k\eta_{init}}{8+6c_s^2-12w_{de}+9c_s^2(w_{de}-c_a^2)} h_{init}. \label{v_de_init}
\end{eqnarray}

More detailed analysis of gravitational stability of the tachyonic field can be found in \cite{Jain2007,Sergijenko2009a,Sergijenko2009b}. For solving the system of differential equations describing the evolution of the scalar cosmological perturbations with adiabatic initial conditions we use the CAMB code, modified for our dark energy model.
 
In fig.~\ref{ddeb} the evolution of amplitudes of dark energy, dark matter and baryons density perturbations with the scale of $k=0.05$ h/Mpc is shown for the model with the most optimal cosmological parameters $\mathbf{q}_1$ for quintessential and $\mathbf{p}_1$ for phantom fields, determined in \cite{Novosyadlyj2012}, with classical and tachyonic Lagrangians. We can see that in the case of quintessential field the difference in evolution of amplitudes of the perturbations of the classical and tachyonic fields is noticeable and we can hope to differ them by observational data. In the case of phantom field we do not see any differences. It can be explained by the fact that in the case of quintessential tachyonic field the square of effective speed of sound $c_s^2$ changes from
0.022 at early epoch to 0.996 at current one, while for the classical field it is constant and equal to 1. In the case of phantom tachyonic field $c_s^2=1$ at the beginning and it grows monotonically to 1.043 in current epoch, which practically does not change the evolution of $\delta_{de}(a)$ in comparison with phantom classical field.

\begin{figure}[tb]
  \centering
  \includegraphics[width=.49\textwidth]{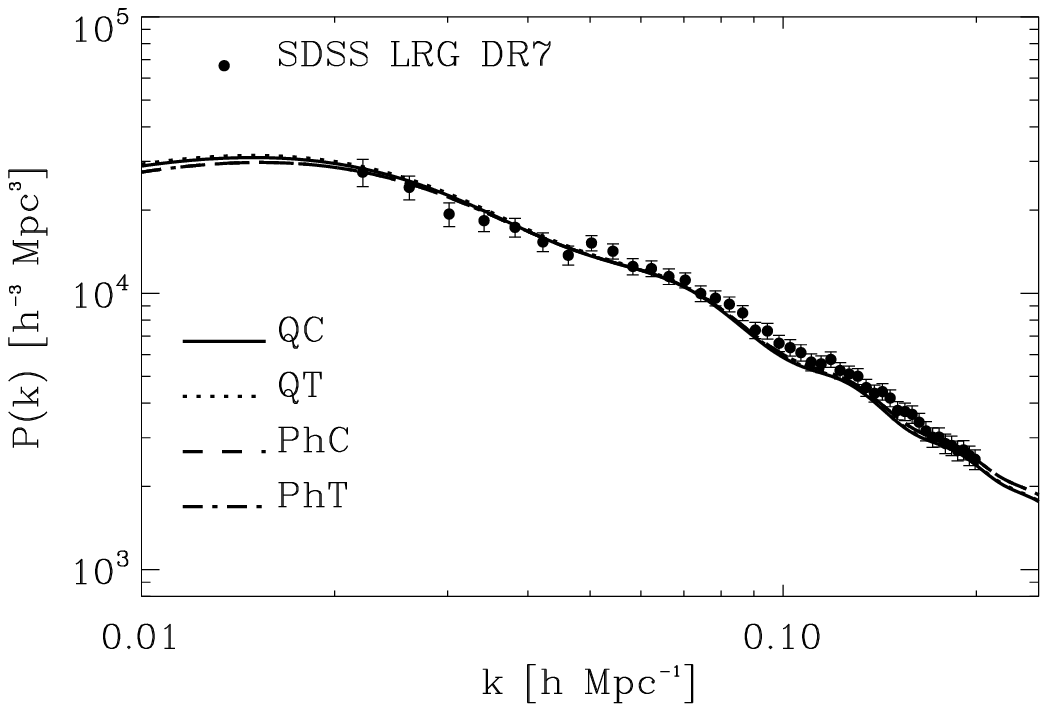}
  \includegraphics[width=.49\textwidth]{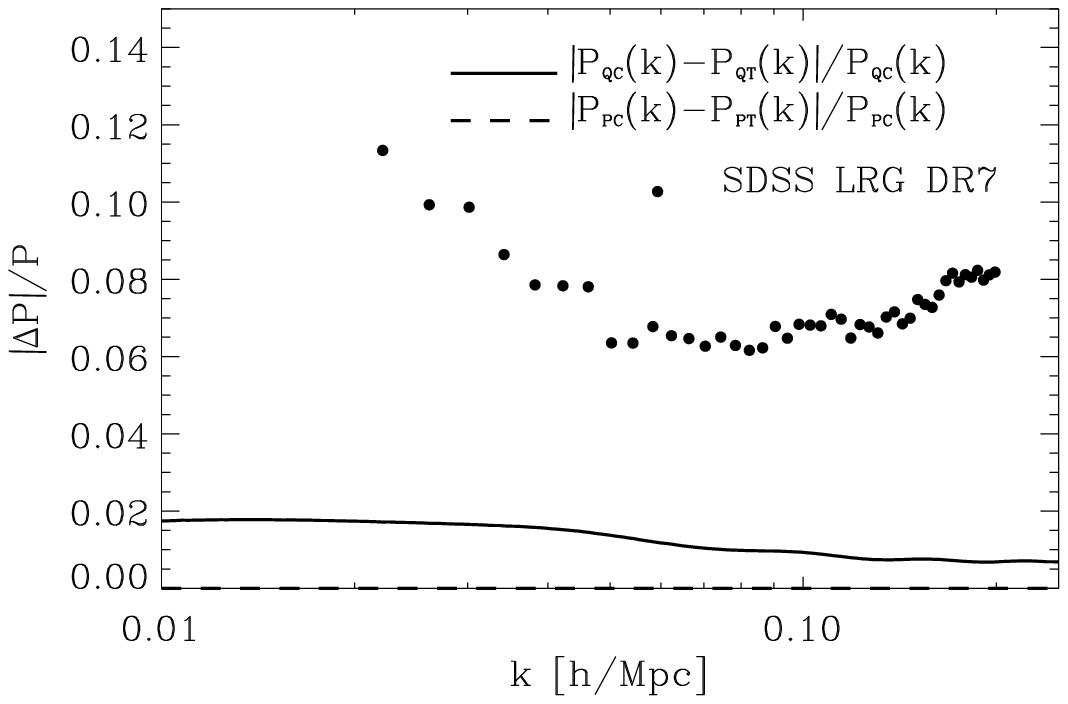}
  \caption{Left: power spectra of the matter density perturbations in the models with quintessential classical and tachyon (QC, QT) and phantom classical and tachyon (PC, PT) scalar fields with the most optimal values of cosmological parameters $\mathbf{q}_1$ and $\mathbf{p}_1$ determined in \cite{Novosyadlyj2012}; circles denote the power spectrum of galaxies spatial inhomogeneities from SDSS LRG DR7~\cite{Reid2010} observational data. Right: relative differences of power spectra in models with the same parameters but different Lagrangians (classical and tachyonic) $|\Delta P(k)|/P(k)$; circles denote the observed uncertainties (1$\sigma$) of the data SDSS LRG DR7~\cite{Reid2010}.}
  \label{pk_qpt}
\end{figure}
In fig.~\ref{pk_qpt} the power spectra of the matter density perturbations for the same models as in fig.~\ref{ddeb} and their differences in the models with quintessential and phantom tachyonic scalar fields are presented. To compare them with observations the data of SDSS LRG DR7~\cite{Reid2010} are shown too. We can see that even in quintessence region this difference does not exceed 2\% and is much less than the observed uncertainties in SDSS LRG DR7 data ($\sim6-12\%$). In phantom region this difference does not exceed $\sim0.001$ of percent (dashed line merges with horizontal axis) and is unreachable for determination by these observational data even in far perspective.

\section{Conclusions}
Scalar fields with the effective Dirac-Born-Infeld Lagrangian describe the tachyonic modes of oscillations of cosmic strings and branes in superstring theories of the fundamental particles, the exponential expansion (inflationary phase) of the very early Universe, moreover, they successfully describe its current accelerated expansion. Tachyon field is indistinguishable from classical scalar field by its cosmological manifestations in model with the same parameters at the level of accuracy of current observational data. In the phantom region of the values of EoS parameter these fields are practically indistinguishable by cosmological data. Their special feature in the case of phantom field is superluminal speed of propagation of perturbations (effective sound speed) with real values of all other physical characteristics of the field. Is it in contradiction with principle of causality, if the tachyon field interacts with other fields and particles only gravitationally? Can this special feature of the field be 
the ground 
for removing tachyon field from the list of candidates for dark energy? The author has no confidently reasoned answers to these questions, but no matter what they are, or what will be the future outcome of establishing the nature of dark energy, the conclusion about fruitfulness of ideas of tachyon and benefit of the discussions of them remains indisputable.

\section*{Acknowledgements}
This work was supported by the project of Ministry of Education and Science of Ukraine (state registration number 
0113U003059) and research program ``Scientific cosmic research'' of the National Academy of Sciences of Ukraine (state registration number 0113U002301). Author also acknowledges the usage of CAMB package.


\begin{thebibliography}{99}
\bibitem{Bilaniuk1962} Bilaniuk O. M. P., Deshpande V. K., Sudarshan E. C. G., ``Meta'' Relativity,
American Journal of Physics \textbf{30}, 718 (1962).
\bibitem{Garousi2000} Garousi M. R.,  Tachyon couplings on non-BPS D-branes and Dirac-Born-Infeld action,
Nucl. Phys. B \textbf{584}, 284 (2000)
\bibitem{Sen2002} Sen A., Tachyon Matter, JHEP \textbf{0207}, 065 (2002).
\bibitem{Garousi2004} Garousi M. R., Sami M., Tsujikawa S., Cosmology from a rolling massive scalar field on the anti-D3 brane of de Sitter vacua, Phys. Rev. D \textbf{70}, 043536 (2004).
\bibitem{Padmanabhan2002} Padmanabhan T., Accelerated expansion of the universe driven by tachyonic matter,
Phys. Rev. D \textbf{66}, 021301 (2002).
\bibitem{Gibbons2002} Gibbons G. W., Cosmological evolution of the rolling tachyon, Phys. Lett. B  \textbf{537}, 1 (2002).  
\bibitem{Frolov2002} Frolov A., Kofman L., Starobinsky A., Prospects and problems of tachyon matter cosmology,
Phys. Lett. B \textbf{545}, 8 (2002).
\bibitem{Bagla2003} Bagla J. S., Jassal H. K., Padmanabhan T., Cosmology with tachyon field as dark energy, Phys. Rev. D
\textbf{67}, 063504 (2003).
\bibitem{Abramo2003} Abramo L. R., Finelli F., Cosmological dynamics of the tachyon with an inverse power-law potential,
Phys. Lett. B \textbf{575}, 165 (2003).
\bibitem{Gibbons2003} Gibbons G. W., Thoughts on tachyon cosmology, Class. and Quant. Grav. \textbf{20}, S321 (2003).
\bibitem{Abramo2004} Abramo L. R., Finelli F., Pereira T. S., Constraining Born-Infeld models of dark energy with CMB
anisotropies, Phys. Rev. D \textbf{70}, 063517 (2004). 
\bibitem{Gorini2004} Gorini V., Kamenshchik A., Moschella U., Pasquier V., Tachyons, scalar fields, and
cosmology, Phys. Rev. D. {\bfseries 69}, 123512 (2004).
\bibitem{Sen2005} Sen A., Remarks on Tachyon Driven Cosmology, Phys. Scripta T \textbf{117}, 70 (2005).
\bibitem{Calcagni2006}  Calcagni G., Liddle A. R., Tachyon dark energy models: Dynamics and constraints, Phys. Rev. D
\textbf{74}, 043528 (2006).
\bibitem{Feinberg1967} Feinberg G., Possibility of Faster-Than-Light Particles, Phys. Rev. \textbf{159}, 1089 (1967).
\bibitem{Perlmutter1998} Perlmutter S., Aldering G., della Valle M. et al., Discovery of a supernova explosion at half the
age of the universe, Nature, \textbf{391}, 51 (1998).
\bibitem{Riess1998} Riess A. G., Filippenko A. V., Challis P. et al., Observational Evidence from Supernovae for an
Accelerating Universe and a Cosmological Constant, Astron. J. 1\textbf{16}, 1009 (1998).
\bibitem{Schmidt1998} Schmidt B. P., Suntzeff N. B., Phillips M. M. et al., The High-Z Supernova Search: Measuring Cosmic
Deceleration and Global Curvature of the Universe Using Type IA Supernovae, Astrophys. J. \textbf{507}, 46 (1998).
\bibitem{Sergijenko2009b} Sergijenko O., Novosyadlyj B., Perturbed dark energy: Classical scalar field versus tachyon,
Phys. Rev. D, \textbf{80}, 083007 (2009).
\bibitem{Babichev2005a} Babichev E., Dokuchaev V., Eroshenko Yu. Dark energy cosmology with generalized linear equation
of state, Classical and Quantum Gravity, \textbf{22}, 143, (2005).
\bibitem{Holman2004} Holman R., Naidu S. Dark Energy from Wet Dark Fluid, arXiv:astro-ph/0408102 (2004).
\bibitem{Sergijenko2009a} Sergijenko O., Kulinich Yu., Novosyadlyj B., Pelykh V., Large-scale structure formation in
cosmology with classical and tachyonic scalar fields, Kinematics and Physics of celestial bodies \textbf{25}, 17 (2009).
\bibitem{Novosyadlyj2010} Novosyadlyj B., Sergijenko O., Apunevych S., Pelykh  V., Properties and uncertainties of scalar
field models of dark energy with barotropic equation of state, Phys. Rev. D \textbf{82}, 103008 (2010).
\bibitem{Novosyadlyj2011} Novosyadlyj B., Sergijenko O., Apunevych S., Distinguishability of scalar field models of dark
energy with time variable equation of state parameter, Journal of Physical Studies \textbf{15}, 1901 (2011).
\bibitem{Novosyadlyj2012} Novosyadlyj B., Sergijenko O., Durrer R., Pelykh V. Do the cosmological observational data prefer
phantom dark energy? Phys. Rev. D \textbf{86} 083008 (2012).
\bibitem{Caldwell2002} Caldwell R. R., A Phantom Menace? Cosmological consequences of a dark energy component with super-negative equation of state, Phys. Lett. B \textbf{545}, 23 (2002).
\bibitem{Babichev2008} Babichev E., Mukhanov V., Vikman A., k-Essence, superluminal propagation, causality and emergent geometry, Journal of High Energy Physics \textbf{02}, 101 (2008).
\bibitem{Ma1995} Ma C.-P. \& Bertschinger E., Cosmological perturbation theory in the synchronous and conformal newtonian
 gauges, Astrophys. J. \textbf{455}, 7 (1995).
\bibitem{Jain2007} Jain R. K., Chingangbam P., Sriramkumar L., On the evolution of tachyonic perturbations at
super-Hubble scales, J. Cosmol. Astropart. Phys. 10, 03 (2007).
\bibitem{Reid2010} Reid B. A., Percival W. J., Eisenstein D. J. et al., Cosmological constraints from the
  clustering of the Sloan Digital Sky Survey DR7 luminous red galaxies, Mon. Not. R. Astron. Soc. \textbf{404}, 60 (2010). 
\end{thebibliography}
\end{document}